# A Simulation Package in VBA to Support Finance Students for Constructing Optimal Portfolios

Abdulnasser Hatemi-J [1*], Alan Mustafa [2]

[1] Department of Economics and Finance, College of Business and Economics, UAE University; AHatemi@uaeu.ac.ae

[2] IEEE, Senior Member, Duhok, Kurdistan Region, Iraq; Alan.Mustafa@ieee.org

\* Correspondence: AHatemi@uaeu.ac.ae

**Abstract:** This paper introduces a software component created in Visual Basic for Applications (VBA) that can be applied for creating an optimal portfolio using two different methods. The first method is the seminal approach of Markowitz that is based on finding budget shares via the minimization of the variance of the underlying portfolio. The second method is developed by El-Khatib and Hatemi-J, which combines risk and return directly in the optimization problem and yields budget shares that lead to maximizing the risk adjusted return of the portfolio. This approach is consistent with the expectation of rational investors since these investors consider both risk and return as the fundamental basis for selection of the investment assets. Our package offers another advantage that is usually neglected in the literature, which is the number of assets that should be included in the portfolio. The common practice is to assume that the number of assets is given exogenously when the portfolio is constructed. However, the current software component constructs all possible combinations and thus the investor can figure out empirically which portfolio is the best one among all portfolios considered. The software is consumer friendly via a graphical user interface. An application is also provided to demonstrate how the software can be used using real-time series data for several assets.

## 1. Introduction

Knowledge delivery as a way of continuation of humanity's mind to the next generation has employed forms of tacit, explicit, and implicit knowledge [1, 2]. One major method which these deliveries are shown through is the teaching and training environment. However, to ensure the delivery of knowledge the learned materials must be assessed. Although continuous investigations are taken place by researchers to seek different ways of compatibility of teaching materials are in-line with learner's learning style. Research has indicated that by personalizing teaching materials to suit the specific needs of a learner, the learning performance has shown improvement [3, 4]. Furthermore, the practice on learned materials emphasizes deep learning and its effect can stay with the learner for a longer time depending on the sessions of practice.

It is widely agreed that not all theories can be directly put onto practice such as aviation (new trainee pilots require many hours of flights before being called a pilot), medical (medical students require long hours of practice in surgery before being allowed to do independent surgery), mathematics (mathematicians and economists require an environment where they can apply theoretical concepts into practice before becoming reality), and manufacturing (which requires a tremendous amount of resources through planning, designing, and implementation of technologies before a tangible product gets produced) and many more industries can be a testimony for showing the value of simulation-based learning can be an advantage on saving resources and money. Also, detailed instruction on the process of solving a problem including given immediate feedback can enhance learning [5, 6, 7, 8].

Hence, in this paper, simulation software has been developed which shows how a decision can be improved before an actual event, such as a decision on portfolio diversification. The following sections describe the logic behind portfolio diversification, mathematical derivation and the rationale behind each method, a chosen group of algorithms for the decision-maker unit (DMU) in both simulation software, and at the end, we discuss our findings, and concluding our work will be presented. The software verbiage is available from the authors on request.

The rest of the paper is disposed as the following. Section 2 describes the two alternative methods that can be used for constructing a portfolio. This section also illustrates how the dimension of a portfolio can be determined endogenously. Section 3 presents our software component and describes how it can be used. Section 4 presents the finding of an empirical application. The last section provides conclusions. In Appendix A, a schematic representation of both modules that are created in this paper is provided.

## 2. Methodology

In this section we described the alternative approaches for constructing financial portfolios. Another issue that is usually neglected in the literature is the dimension of the portfolio. It is a common practice in literature to assume that the number of assets included in the portfolio is given *a priory*. However, this does not need to be the case in real markets. For many investors the selection of assets is also an endogenous question. Our software takes this issue into account by constructing all possible combinations and providing the portfolio that is optimal even with regards to the number of assets also. This approach is described in the sections sub-section.

*2.1 Portfolio Construction*

The seminal method for portfolio diversification is established by Markowitz [9], which leads to obtaining budget shares via minimizing the variance of the selected portfolio with regards to the budget restriction. Let us assume that $r_i$ is representing the rate of return for asset i, which is has a normal as $r_i \sim \Phi(\bar{r}_i, \sigma_i^2)$. The variance and covariance matrix for the assets included in the portfolio (denoted by n) is expressed as $\Omega = (\sigma_{i,j})_{1 \leq i,j \leq n}$, here $\sigma_{ij}$ is the covariance measure between the returns of the two assets i and j. Let us also define $w_i$ as the weight for asset i. Therefore, the average return of the portfolio is defined as $F(w) = \sum_{i=1}^{n} \bar{r}_i w_i$ and its variance as a measure of risk is $V(w) = w'\Omega w$. Hence, the optimization objective of Markowitz [9] is the following minimization problem:

$$Minimize\ V(w) = w'\Omega w \tag{1}$$

Bounded by the budget limitation expressed in equation (2):

$$D(w) = \sum_{i=1}^{n} w_i = 1 \tag{2}$$

The solution for each $w_i$ of this optimization problem is obtained as the following, assuming that there are four assets in the portfolio (i.e., *n=4*):

$$w_1 = -\frac{\begin{vmatrix} B_{1,2} & B_{1,3} & B_{1,4} \\ B_{2,2} & B_{2,3} & B_{2,4} \\ B_{3,2} & B_{3,3} & B_{3,4} \end{vmatrix}}{|E|} \tag{3}$$

$$w_2 = \frac{\begin{vmatrix} B_{1,1} & B_{1,3} & B_{1,4} \\ B_{2,1} & B_{2,3} & B_{2,4} \\ B_{3,1} & B_{3,3} & B_{3,4} \end{vmatrix}}{|E|} \tag{4}$$

$$w_3 = -\frac{\begin{vmatrix} B_{1,1} & B_{1,2} & B_{1,4} \\ B_{2,1} & B_{2,2} & B_{2,4} \\ B_{3,1} & B_{3,2} & B_{3,4} \end{vmatrix}}{|E|} \tag{5}$$

$$w_4 = \frac{\begin{vmatrix} B_{1,1} & B_{1,2} & B_{1,3} \\ B_{2,1} & B_{2,2} & B_{2,3} \\ B_{3,1} & B_{3,2} & B_{3,3} \end{vmatrix}}{|E|} \quad (6)$$

where

$$E = \begin{pmatrix} B_{1,1} & B_{1,2} & B_{1,3} & B_{1,4} \\ B_{2,1} & B_{2,2} & B_{2,3} & B_{2,4} \\ B_{3,1} & B_{3,2} & B_{3,3} & B_{3,4} \\ 1 & 1 & 1 & 1 \end{pmatrix}$$

Observe that $|Y|$ signifies the determinant of the matrix $Y$. Also notice that $B$ is an $n \times n$ matrix, which has the following attributes:

$$B_{i,j} = (\sigma_{i+1,j} + \sigma_{j,i+1}) - (\sigma_{i,j} + \sigma_{j,i}), \quad \forall \quad 1 \leq i \leq n-1 \quad and \quad \forall \quad 1 \leq j \leq n.$$

The Markowitz approach, which is commonly utilized by investors, constructs a portfolio that has the smallest possible risk. Nonetheless, it is broadly agreed that the rational investors pay attention to both risk and return when investment decisions are made. Consequently, Hatemi-J and El-Khatib [10] propose optimizing the portfolio diversification problem that combines risk and return directly when the portfolio is created. Specifically, the objective function in the optimization problem is the following as per the authors:

$$Maximize \; \frac{F(w)}{\sqrt{V(w)}} = \frac{F(w)}{\sqrt{w'\Omega w}} \quad (7)$$

subject to

$$D(w) = \sum_{i=1}^{n} w_i = 1 \quad (8)$$

Via the application of Theorem 1 in Hatemi-J, Hajji and El-Khatib-[11], the solutions for the optimal budget shares within this setting are provided in the following equations, when $n=4$:

$$w_1 = \frac{-\begin{vmatrix} G_{1,2} & G_{1,3} & G_{1,4} \\ G_{2,2} & G_{2,3} & G_{2,4} \\ G_{3,2} & G_{3,3} & G_{3,4} \end{vmatrix}}{|K|} \quad (9)$$

$$w_2 = \frac{\begin{vmatrix} G_{1,1} & G_{1,3} & G_{1,4} \\ G_{2,1} & G_{2,3} & G_{2,4} \\ G_{3,1} & G_{3,3} & G_{3,4} \end{vmatrix}}{|K|} \quad (10)$$

$$w_3 = \frac{-\begin{vmatrix} G_{1,1} & G_{1,2} & G_{1,4} \\ G_{2,1} & G_{2,2} & G_{2,4} \\ G_{3,1} & G_{3,2} & G_{3,4} \end{vmatrix}}{|K|} \tag{11}$$

$$w_4 = \frac{\begin{vmatrix} G_{1,1} & G_{1,2} & G_{1,3} \\ G_{2,1} & G_{2,2} & G_{2,3} \\ G_{3,1} & G_{3,2} & G_{3,3} \end{vmatrix}}{|K|} \tag{12}$$

Where

$$K = \begin{pmatrix} G_{1,1} & G_{1,2} & G_{1,3} & G_{1,4} \\ G_{2,1} & G_{2,2} & G_{2,3} & G_{2,4} \\ G_{3,1} & G_{3,2} & G_{3,3} & G_{3,4} \\ 1 & 1 & 1 & 1 \end{pmatrix}$$

Observe that *G* is an *n*×*n* matrix that has the following definition:

$$G_{i,j} = \bar{r}_i(\sigma_{i+1,j} + \sigma_{j,i+1}) - \bar{r}_{i+1}(\sigma_{i,j} + \sigma_{j,i}), \quad \forall \ 1 \le i \le n-1 \quad and \quad \forall \ 1 \le j \le n.$$

Accordingly, this new method merges risk and return in the optimization problem, which accords well with reality. This is the case because rational investors consider both risk and return when they make any investment decision.

*2.2.The Dimension of a Portfolio*

Prior to finding the budget shares, the investor must choose the assets to include in the portfolio. This is a crucial matter. The way the literature is dealing with this issue is to assume that the number of assets included in the portfolio is provided exogenously. Nevertheless, this is not the way the investors approach in real markets. The selection of assets for including in the portfolio is better dealt with as an endogenous question according to Hatemi-J and Hajji [12]. The authors suggest a solution, which is based on selecting the maximum number of assets that the investor might be interested in based on his/her subjective preferences. Subsequently, a series of portfolios containing different permutations of these assets can be created. Suppose that *n* is the maximum number of the assets that is considered by the investor for potentially including in the portfolio. Thus, the number of combinations (denoted by *P*) needs to be built is the following according to Hatemi-J and Hajji [12]:

$$P = \sum_{l=0}^{n-2} C(n, n-l) = \sum_{l=0}^{n-2} \frac{n!}{(n-l)! \times l!} \tag{13}$$

Therefore, $P$ is the total number of permutations that are accessible to the investor as portfolios for a given $n$ set of underlying assets. By creating all these $P$ portfolios, the investor should calculate the risk adjusted return for each portfolio in this set. For instance, when $n$ is 4 then $P$ is equal to 11 portfolios based on equation (13) as it is the case in our application. Via the risk adjusted returns for these $P$ portfolios, the investor can retrieve the best portfolio, the second best, or the third best etc. This approach makes it operational to obtain the portfolio amongst these 11 portfolios that produces the highest magnitude of return for each unit of risk. That is, the best portfolio ($BP$) among the $P$ combinations is acquired as

$$BP = Max[RAR_k, \cdots, RAR_P] \qquad (14)$$

Where

$$RAR_k = \frac{E[R_{pk}(w)]}{\sqrt{V[R_{pk}(w)]}} \qquad (15)$$

The denotation $E[R_{pk}(w)]$ represents the expected return of portfolio $k$ (for $k = 1, \ldots, P$) for the given vector for optimal budget shares (i.e., $w$). $V[R_{pk}(w)]$ represents the variance of the same portfolio and $RAR_k$ denotes the risk adjusted return of portfolio $k$. The needed optimal budget shares for each portfolio might be acquired via minimizing the variance of the portfolio as established by Markowitz [9]. Nonetheless, it is also feasible to find the optimal budget shares via the method introduced by Hatemi-J and El-Khatib [10] and generalized by Hatemi-J, Hajji and El-Khatib [11]. These methods are described by equations (1-12) above.

## 3. Experimental Design (designing a new tool)

In this section, designing a new tool to present the process of simplifying and solving a complicated process which usually takes a long time using either pen and paper, or calculators, or a simple spreadsheet for manually doing the calculations has been recommended. Authors have used the power of Microsoft Visual Basic for Applications (VBA) in Microsoft Excel to create this module to automate lengthy processes of creating multiple portfolios and their comparison to select the best possible choice of a portfolio from a list of diversified portfolios, and of course the efficiency of the work is incomparable. Since the use of VBA as a tool to automate complex calculations in the industry has become a norm. Kalwar et. al., [13] has given a productive list of VBA applications in the industry which clearly backs its statement. Blayney et al. [14] also present the capabilities and use of VBA in conjunction with MS Excel to do preliminary analysis in big data research.

As an example, one of the commonly used methods in finance is portfolio diversification. A personal investor or a financial organization's task is to do investment on a series of instruments. Those assets could be chosen from any of commodities, indices, forex, metal, energy, and stocks as a few to name. The issue here is what would be the best combination of those assets for the investment based on their historical market price and by minimizing the risk involved in trading those markets.

Markowitz [9] in his paper recommended a solution for finding the optimal selection on the best combination of assets for investment. The approach was mainly based on the weights as budget shares that minimized the variance of the underlying portfolio. In this approach, however, the risk on the amount of return has not been considered. Hatemi-J and El-Khatib [10] devised a new method based on the effects of valuing risks on the selection of assets so to maximize the return. The method is named "maximizing the risk-adjusted return of the portfolio". It combines risk and returns when the optimal budget shares are searched for. Hence, two applications are presented in this section. The basis of the design for the first application is a set of a predetermined list of assets (denoted by PD-RAR which stands for Portfolio Diversification with Risk Adjusted Return), and the second one is presenting a comparison between portfolios based on different number of assets (denoted PD-RAR-Comb). This last design is aimed at helping the investor to endogenize the number of assets in the portfolio by considering all possible combinations. Equations 1-14 are used for this purpose.

## 4. Development of the Tool

Since there are complex calculations involved in calculating portfolios with the best performance on both methods of Markowitz [9] and Hatemi-J and El-Khatib [10], there is a need for developing a module that does all the required calculations efficiently via Graphical User Interfaces (GUI). The aim of this research is to fill this gap in the existing literature. Schematics of these two designs as presented in the previous section are given in Appendix A.

The portfolio diversification with risk adjusted return (PD-RAR) creates a portfolio for a set of data input via its dashboard panel (Figure 1). There are two methods available for entering data: i) either through copy and paste functions to paste the set of data on the sheet named 'Data', or ii) the option named 'Data as Parameters'. The use of the first option is straight forward, and the application is ready to process data. The second option provides an extra option for entering the input data in the form of available number of assets, calculated expected values, and covariance of the set of data, which all could have come from another application software (Figure 2).

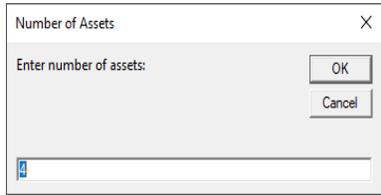

Figure 1: The main dashboard panel

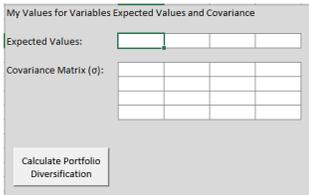

(a)                                                                                  (b)

Figure 2: By following the option of 'Data as Parameters' two dialog boxes are presented; **(a)** entering number of assets, **(b)** a dialog box ready to enter values for average returns, and the variance-covariance.

After processing data based on the equations 1-12, the number of portfolios is given (Figure 3). It follows up with creating two types of output: i) detailed calculations of applied algorithms (Figure 4), and ii) the ''Estimated Results" (Table 1 and Table 2).

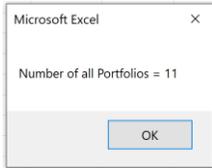

Figure 3: The notification dialog box shows the possible number of portfolios created.

[Figure 4]: Sample of detailed calculations based on equations 1-12

Table 1: Summary of estimation results table for portfolio number 10 with best value based on minimum variance approach (MV), which includes two assets.

| Assets | Average Return ($\bar{r}$) | SD ($\sigma$) | Risk Adjusted Return | w for MV | w for MRAR |
|---|---|---|---|---|---|
| **Brent Oil** | 0.00364822 | 0.01435503 | 0.25414215 | 0.17252711 | 0.4888804 |
| Dow Jones | 0.0017301 | 0.00862258 | 0.20064716 | 0.82747289 | 0.5111196 |
| Portfolio - Minimum Variance (MV) | 0.00206103 | 0.00826817 | 0.24927208 | | |
| **Portfolio - Maximum Risk Adjusted Return (MRAR)** | 0.00266783 | 0.00940691 | 0.28360313 | | |

**Table 2:** Summary of estimation results table for portfolio number 1 with the best value based on maximum risk-adjusted return (MRAR), which includes all assets.

| Assets | Average Return (r̄) | SD (σ) | Risk Adjusted Return | w for MV | w for MRAR |
|---|---|---|---|---|---|
| USD-JPY | 0.00029673 | 0.00349409 | 0.0849228 | 1.09569014 | 0.11642855 |
| Brent Oil | 0.00364822 | 0.01435503 | 0.25414215 | 0.08033079 | 0.46561635 |
| DAX | 0.00142506 | 0.00937355 | 0.15203045 | -0.07538021 | -0.11039877 |
| Dow Jones | 0.0017301 | 0.00862258 | 0.20064716 | -0.10064071 | 0.52835387 |
| Portfolio - Minimum Variance (MV) | 0.00033665 | 0.00321823 | 0.10460585 | | |
| Portfolio - Maximum Risk Adjusted Return (MRAR) | 0.00249 | 0.00875246 | 0.284491 | | |

The calculation results are provided in Table 3 for the best portfolio that is created based on maximum risk-adjusted return.

**Table 3:** List of Portfolios with the Minimum Variance (MV) and Maximum Risk-Adjusted Return (MRAR) with the highest selection.

| Portfolio | MV | MRAR | Portfolio with the highest RAR [1] | |
|---|---|---|---|---|
| Portfolio - 1 | 0.10460585 | 0.284491 | MV | MRAR |
| Portfolio - 2 | 0.12222364 | 0.26876603 | 0.24927208 | 0.284491 |
| Portfolio - 3 | 0.10252419 | 0.28364519 | Portfolio 10 | Portfolio 1 |
| Portfolio - 4 | 0.0426772 | 0.20488306 | | |
| Portfolio - 5 | 0.23869796 | 0.28430838 | | |
| Portfolio - 6 | 0.13814161 | 0.26645813 | | |
| Portfolio - 7 | 0.0582005 | 0.15205724 | | |
| Portfolio - 8 | 0.04421402 | 0.20429712 | | |
| Portfolio - 9 | 0.21516302 | 0.26479205 | | |

| | | |
|---|---|---|
| Portfolio - 10 | 0.24927208 | 0.28360313 |
| Portfolio - 11 | 0.19537426 | 0.20076821 |

[1] Portfolio Construction Methods: MV= Minimum Variance Approach; MRAR = Maximum Risk Adjusted Return Approach.

*3.2. PD-RAR-Comb*

In this sub-section the results for all eleven combinations are briefly presented in Table 3. These results are obtained by using the PD-RAR-Comb module. This module provides a list of all possible combinations of portfolios to be created for all assets with the addition of presenting a comparison between both algorithms used in creating those portfolios [9], [10] [Appendix A.2]. The first phase of this process is to create a list of possible combinations of assets, then creating a portfolio for each combination, and list them in a sheet with the arrangement descending order of number of assets, from maximum number of assets to their minimum number of assets in a combination. Next step is to find the maximum value for both used algorithms of minimum variance (MV) and maximum risk-adjusted return (MRAR). The outcome of this process is shown in Table 3.

## 5. Findings

In this section, the findings for executing both modules are discussed. Although, the performance of calculations mainly depends on the type of data processing (i.e., the option of 'with or without detailed presentation of step-by-step calculations'), and on the size of the dataset (i.e., number of assets and records closing prices for each asset). For example, by running a set of 10 assets with 65 records (that results in 1013 different portfolios), it takes around 100 seconds to process the data using the option "without details". Comparatively, it takes more than 15 minutes to implement the same calculations with the option "with details". The reason for this is due to the interaction with an individual worksheet (i.e., reading and writing data from and into a worksheet). It should be mentioned that the main purpose of using the module with detailed steps is for educational purposes, which gives the outcome of step-by-step calculations in the process of creating portfolios.

Note that the portfolio construction based on Hatemi-J and El-Khatib [4] method clearly shows better outcome if the goal is finding a portfolio that provides the highest possible return per unit of risk. However, the portfolio that is constructed by Markowitz's [9] method results in the lowest possible risk. By using this module, it is also possible to directly enter the parameters that are necessary inputs for portfolio diversification (such as the average returns and the variance-covariance values) instead of importing the time-series data of the prices (see Figure 2).

## 6. Conclusions

Constructing an optimal portfolio is an important issue for investors and financial institutions. There are several methods available in literature for this purpose. The seminal approach provided by Markowitz [9] yields an optimal portfolio that results in the minimum possible risk. This approach is widely applied by practitioners. An alternative approach that is developed by Hatemi-J and El-Khatib [10], produces a conditional optimal portfolio that gives the maximum return per unit of risk. The aim of this work is to provide a VBA module that can construct portfolios using both methods.

A pertinent issue within this context, which is usually neglected in the literature, is the dimension of the portfolio. That is, the number of assets included in the portfolio is assumed to be exogenous. However, it is rational to deal with this issue endogenously. The approach that is suggested by Hatemi-J and Hajji [12] for this purpose is to estimate all possible combinations of portfolios and estimate the risk adjusted return for each. The portfolio that gives the highest risk adjusted return among all possible ones is the one that should be selected. Our module provides this possibility also. It constructs all possible portfolios that the investor might be interested in and indicates the optimal one using both portfolio diversification methods. An example of four assets is provided to demonstrate how the module operates. However, the results can be generalized in future applications. The module is very consumer-friendly indeed. The software verbiage of the module is accessible from the authors on demand.

**Appendix A:**

Dataflow and schematics for system design of both recommended modules of PD-RAR and PR-RAR-Comb.

**A.1:** Figure A.1 presents a data processing mechanism for calculating the PD-RAR. Detailed mathematical algorithm for this process is given in equations 1-12 in section 2.1, and a screenshot of its output is given in Table 1 and Table 2.

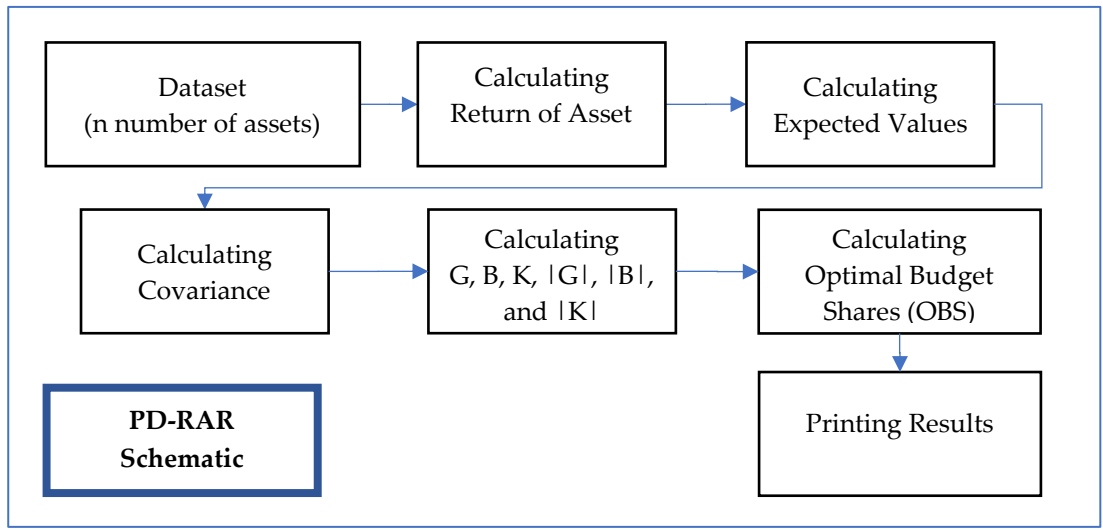

Figure A.1: Schematic diagram of portfolio diversification for PD-RAR model

**A.2:** Figure A.2 gives a list of combinations of portfolios developed based on i) maximum values of minimum variance (MV), and ii) maximum risk adjusted return approaches (MRAR). The process is using the same mechanism of calculating portfolios with an addition of a looping through portfolios. The outcome is given in Table 3.

Based on equations 1-15 of sections 2.1 and 2.2, for the implantation of the module.

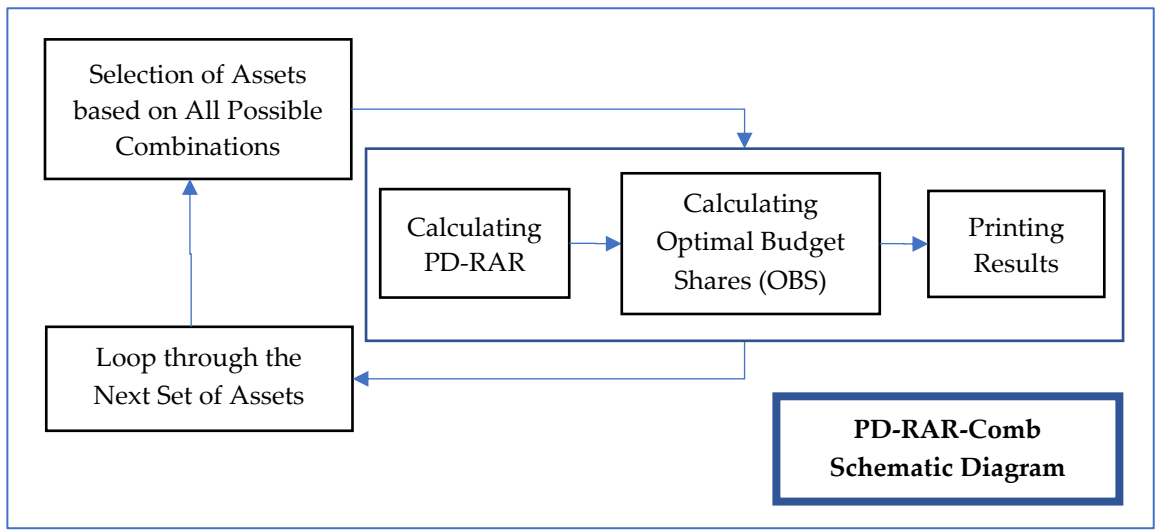

Figure A.2: Schematic diagram of portfolio diversification for PD-RAR-comb model